\documentclass[epj]{svjour}
\usepackage[utf8]{inputenc}    
\usepackage[T1]{fontenc}       
\usepackage{amsmath}
\usepackage{graphics}
\usepackage[dvipsnames]{xcolor}
\usepackage{hyperref}
\hypersetup{colorlinks,allcolors=black}
\usepackage{graphicx}
\usepackage{caption,subcaption}
\makeatletter
\renewcommand{\fnum@figure}{Figure \thefigure}
\makeatother
\usepackage{tikz}
\usetikzlibrary{calc,matrix,positioning}
\usetikzlibrary{decorations.pathmorphing,patterns}
\tikzset{
declare function={
        f(\z,\v,\x) = \z+\v*\x-0.5*\g*\x^2;
        }
}
\usepackage{pgfplots}
\pgfplotsset{
    ,compat=newest
    }
\usepgfplotslibrary{groupplots}
\usepgflibrary{patterns}
\usepackage{pgfplotstable}
\pgfplotstableread{./SECTIONS/Data/DensityDataBEC.txt}{\DensityDataBEC}
\pgfplotstableread{./SECTIONS/Data/DensityDataTHE.txt}{\DensityDataTHE}
\pgfplotstableread{./SECTIONS/Data/ErrorDataBEC.txt}{\ErrorDataBEC}
\pgfplotstableread{./SECTIONS/Data/ErrorDataTHE.txt}{\ErrorDataTHE}%
\newcommand{\dataBECdkc}{./SECTIONS/Data/dkc_BEC.dat}
\newcommand{\dataBECnodkc}{./SECTIONS/Data/nodkc_BEC.dat}
\newcommand{\dataCLSdkc}{./SECTIONS/Data/dkc_CLS.dat}
\newcommand{\dataCLSnodkc}{./SECTIONS/Data/nodkc_CLS.dat}
\usepackage{multirow}
\usepackage{lscape}
\usepackage{rotating}
\usepackage{tabularx}
\usepackage{booktabs}
\begin{document}
\title{Precision inertial sensing with quantum gases}
\subtitle{A comparative performance study of condensed versus thermal sources for atom interferometry}
\author{T. Hensel\inst{1}$^{,}$\inst{2}, S. Loriani\inst{1}, C. Schubert\inst{3,1}, F. Fitzek\inst{1}, S. Abend\inst{1}, H. Ahlers\inst{1}, J.-N. Siem\ss\inst{2}$^{,}$\inst{1}, K. Hammerer\inst{2}, E. M. Rasel\inst{1} and N. Gaaloul\inst{1}}
\authorrunning{T. Hensel et al.}
\titlerunning{Precision inertial sensing with quantum gases}
\offprints{\href{mailto:gaaloul@iqo.uni-hannover.de}{gaaloul@iqo.uni-hannover.de}}
\institute{
Institute of Quantum Optics, Leibniz University Hannover, Welfengarten 1, 30167 Hannover
\and Institute for Theoretical Physics, Leibniz University Hannover, Appelstraße 2, 30167 Hannover
\and Deutsches Zentrum für Luft- und Raumfahrt e.V. (DLR), Institut für Satellitengeodäsie und Inertialsensorik, c/o Leibniz Universität Hannover, DLR-SI, Callinstraße 36, 30167 Hannover, Germany
}
%
%
\abstract{
Quantum sensors based on light-pulse atom interferometers allow for high-precision measurements of inertial and electromagnetic forces such as the accurate determination of fundamental constants as the fine structure constant or testing foundational laws of modern physics as the equivalence principle. These schemes unfold their full performance when large interrogation times and/or large momentum transfer can be implemented. In this article, we demonstrate how precision interferometry can benefit from the use of Bose-Einstein condensed sources when the state of the art is challenged. We contrast systematic and statistical effects induced by Bose-Einstein condensed sources with thermal sources in three exemplary science cases of Earth- and space-based sensors.
}
\maketitle
\section{Introduction}
\label{intro}
Atom interferometers are mainly used for inertially sensitive measurements~\cite{Geiger2020arxiv} and a variety of tests of fun\-da\-men\-tal physics~\cite{Bouchendira2011,Asenbaum2020arxiv}.
Key levers to increase the sensitivity are the transfer of a large number of photons during the beam-splitting processes, extending the time of free fall while maintaining contrast and atomic flux.
At the same time, the characterization of errors requires an increased level of control over the manipulation and preparation of atoms.
Limitations of interferometers operating with molasses-cooled atoms and their mitigation by reducing the residual expansion rates were studied theoretically~\cite{Loriani2019} and experimentally~\cite{LouchetChauvet2011,Schkolnik2015}.

In this paper, together with the trade-off between flux and expansion rate, we contrast the appropriateness of the two regimes of atomic ensembles to perform precision tests by evaluating the respective shot noise, cycle times, excitation rates and most prominent systematics such as gravity gradients (GGs), Coriolis force, wave-front aberrations (WFA) and mean-field interactions.

Proposals for space missions, in particular, rely on Delta-Kick collimation (DKC) via optical or magnetic potentials to exploit extended times of free fall in microgravity~\cite{Chu1986,Ammann1997,Mntinga2013,Kovachy2015} and achieve extremely low wave packet expansion rates, corresponding to pK temperatures in thermal ensembles.
Bose-Einstein-condensed (BEC) ensembles are better suited for DKC aiming at long interrogation times~\cite{Loriani2019}, but seem to suffer from a reduced atomic flux due to the evaporation despite recent promising studies~\cite{Rudolph2015}.
Molasses-cooled atoms feature a higher number of atoms, but are typically velocity-filtered in 1D~\cite{Kasevich1991}, which ultimately implies a lower flux of atoms as we will detail in our study.

To illustrate our comparative study between condensed and thermal sources, we consider three prominent cases for free-fall atom interferometers: a gravimeter, a gravity-gradiometer, and a test of the universality of free fall (also known as Weak Equivalence Principle (WEP) test). Thermal sources are defined, in this study, as atomic ensembles with a vanishing condensed fraction. Conversely, BEC sources possess a condensed fraction of $100\%$.
For each case, we limit the maximum allowed diameter of the atomic ensemble at the recombination pulse to preserve contrast.
Subsequently, this enables the determination of the shot noise and other error terms for the trade-off.

This article is structured as follows. Starting with a brief overview of the state of the art of light-pulse atom interferometry, we continue with relevant error contributions to the read-out phases of atom interferometers, quantitatively evaluating them in the three study cases and - finally - discussing the limits of the condensed or thermal regime of the respective interferometry source.
\subsection{State of the art}
Experiments based on the interference of freely falling atoms measure accelerations~\cite{Peters1999,Geiger2020arxiv}, rotations~\cite{Gustavson1997,Gustavson2000,Durfee2006,Canuel2006,Dutta2016}, gravity gradients~\cite{Snadden1998,Bertoldi2006}, determine fundamental constants~\cite{Bouchendira2011,Parker2018,Rosi2014}, perform tests of fundamental physics~\cite{Fray2004,Bouchendira2011,Rosi2014,Schlippert2014,Tarallo2014,Zhou2015,Asenbaum2020arxiv,Albers2020}, and are proposed for the detection of gravitational waves~\cite{Graham2013,Hogan2016,Loriani2019,Schubert2019arxiv,Canuel2018}.
A recent review of the advances in the field of inertial sensing collects most relevant experiments and proposals so far~\cite{Geiger2020arxiv}.
Beyond proof-of-principle demonstrations, ongoing developments target commercialisation as well as challenge the state of the art in sensor performance and in fundamental science.

The most prominent examples of atomic inertial sensors with thermal ensembles are gravimeters~\cite{Freier2016,Mnoret2018} that reach for example a sensitivity of $1.4\times 10^{-8}$\,g at 1\,s~\cite{LeGout2008} and gyroscopes, which measure rotations below $10^{-10}$\,rad/s in a few 100\,s~\cite{Dutta2016}.
A compact prototype  of a gravimeter using BECs reaches a sensitivity of $\delta g/g=3.7\times10^ {-6}$ per cycle~\cite{Abend2016}. This suggests the possibility of a targeted stability on the order of $7.8\times10^{-10}$ after 100\,s of integration time with a state of the art BEC source~\cite{Rudolph2015}.
In a proof-of-principle experiment ~\cite{Dickerson2013}, a point-like BEC source for atom interferometers is implemented in a large fountain experiment to achieve sensitivities of $6.7\times10^{-12}$\,g.

The gravitational constant $G$ is determined to a value $G=6.67191(99)\times 10^{-11}$\,m$^3$kg$^{-1}$s$^{-2}$ with a relative uncertainty of 150\,ppm~\cite{Rosi2014} limited by the initial velocity spread of the atoms in the interferometer.

Most recently~\cite{Parker2018,Yu2019,Clad2019}, there has been extensive work on the determination of the fine-structure constant $\alpha$ via determination of the ratio $\hbar/m$ with matter-wave interferometry, where~$m$ is the atomic mass and~$\hbar$ is Planck's reduced constant.
In a fountain with thermal cesium atoms, the fine-structure constant is determined with an expected statistical error of 0.008\,ppb~\cite{Yu2019}.
The systematics are at the 0.12\,ppb level, mainly stemming from spurious accelerations.
With thermal rubidium, $\hbar/m$ is measured at the $5\times 10^{-9}$ level~\cite{Clad2019}.
An ytterbium contrast interferometer with BECs~\cite{Jamison2014} is used to demonstrate an $\hbar/m$-measurement using large momentum transfer, controlling diffraction effects and atomic interactions with suppression of vibrational effects allowing sub-ppb precision.

In \cite{Zhou2015}, a dual-species WEP test with $^{85}$Rb and $^{87}$Rb reaches a statistical uncertainty of $\eta=0.8\times 10^{-8}$ and is limited by systematic effects, e.g. the Coriolis effect to $\eta=(2.8\pm 3.0)\times 10^{-8}$. Most recently, this limit has been pushed further down to $\eta=1.6\pm5.2\times 10^{-12}$ \cite{Asenbaum2020arxiv}.
The STE-QUEST mission~\cite{Aguilera2014} aims at testing the WEP at the $10^{-15}$ level and beyond by measuring the differential acceleration of a $^{87}$Rb BEC and a $^{41}$K BEC over a total mission time of 5 years~\cite{Battelier2019}.
The concept Quantum Test of the Equivalence principle and Space Time (QTEST)~\cite{Williams2016} intends to determine $\eta$ at the $10^{-15}$ level with thermal ensembles over four integration periods of three months each aboard the International Space Station. 
\subsection{Performance indicators}
In the previous subsection, the state-of-the-art sensitivities for measurements of rotations, accelerations, the fine-structure constant and the Eötvös ratio have been stated. The phase that is to be determined depends on several experimental parameters like the effective wave vector $k_\text{eff}$, the interrogation time 2T, the velocity $v$ of the atomic ensemble perpendicular to the sensitive axis, the length of the detector baseline $L$ or $D$ and the frequency $f$ of the gravitational wave. For the commonly-proposed interferometry schemes discussed above, one can summarize the performance-defining scaling factors to be:
    \begin{itemize}
        \item $k_\text{eff}T^2$ for gravimetry, WEP tests and $G$ measurements,
        
        \item $k_\text{eff}^2T$ for $\hbar/m$ measurements,
        
        \item $k_\text{eff}L\cos(f T)$ for gravitational wave detection,
        
        \item $k_\text{eff}DT^2$ for gravity gradiometry,
        
        \item $k_\text{eff}T^2v$ for rotations.
    \end{itemize}
Increasing these scaling factors allows an improvement in sensitivity.

In this paper we consider shot-noise-limited measurements, where the single-shot phase uncertainty is given by 
\begin{equation}
    \sigma_{\phi_\text{SN}}=1/(C\sqrt{N_\text{at}}),
\end{equation}
defined by the number of interfering atoms $N_\text{at}$ and interferometric contrast $C$. The experiments are repeated n$_\text{cycle}$ times to average (`integrate') the noise of a single-shot phase.

With the assumption of a shot-noise-limited measurement, a fixed atom number and no reduction in contrast $C$, an increase in the scale factor by enhancing the free evolution time $T$ or the effective wave number $k_\text{eff}$ can increase the single-shot phase sensitivity.
Interrogation times 2T of several seconds were realized~\cite{Dickerson2013} and an extension to 10\,s was proposed on space platforms~\cite{Aguilera2014}. The effective momentum transfer ranges from few 10 k$_\text{eff}$ for a single multi-photon pulse up to a few 100\,s of k$_\text{eff}$~\cite{Gebbe2019arxiv} for benchmark experiments.
The integration time to reach the desired performance may range from typically $10^4$\,s up to several months.
Generally, the large number of atoms in thermal ensembles is an advantage over BECs to reduce shot noise. Their increased spatial extension is expected to suppress mean-field effects efficiently compared to BECs. On the other hand, the exact same position and velocity distributions might limit the scaling factors that one could achieve due to large systematic uncertainties and atom losses. In the next section, these systematics and other potentially performance-limiting effects are quantified.
\section{Performance limiting effects in atom interferometry}
\label{sec:physics}
Apart from shot noise considerations, a variety of physical phenomena constitute limiting factors for precision experiments by coupling to the velocity spread or the spatial extension of the ensemble, as is the case for the Coriolis effect, GGs, WFA or mean-field effects. In the following, we characterize different systematic and statistical effects that might limit near-future experiments beyond state of the art such as long-fountain atomic gravimeters, space-based atom interferometers and atom interferometers operated in ground-based laboratories in micro-gravity environments. Starting with intrinsic loss mechanisms due to matter-light interaction, we go on to discuss DKC as a technique to suppress sys\-te\-ma\-tic effects. An analysis of the effects of imperfect detection of the atomic sample concludes this section.
\subsection{Coherent manipulation}
The fidelity of the interferometric beam splitters and mirrors realized by the coherent manipulation of the atoms using light is closely connected to the phase-space properties of the atomic ensemble.

First, homogeneous excitation of the atomic ensemble requires a constant Rabi frequency over the spatial extent of the atoms, which in turn implies a laser beam size much larger than the ensemble size. 
In cases of optical power constraints, e.g. typical for space missions, the ensemble size is hence restricted in order to maintain contrast by achieving reasonable rates for coherent manipulation. 
For large free-fall times such a requirement can be translated into a maximum expansion rate of the ensemble. 
\autoref{fig:DKC_comparison} shows the significant difference in expansion rate between thermal and condensed ensembles which indicates a clear advantage of the latter especially for large interferometry time scales.
Second, for all applications outlined in the previous section, typically Doppler-sensitive two-photon couplings are employed, such that the longitudinal atomic velocity introduces a detuning. 
The temporal profile of light pulses determines their velocity acceptance and hence defines criteria for the atoms' velocity dispersion. 
In general, smaller velocity widths, corresponding to a smaller distribution of detunings, are desirable for an efficient addressing of the atoms.

An effective, simplified model~\cite{CheinetPhD} allows to assess the effects of spatial and velocity selectivity quantitatively. 
The convolution of the atoms' radial density distribution $n(\vec r,t)$  and longitudinal dispersion $f(\vec v)$ with the position- and velocity-dependent excitation rate of the pulse
\begin{equation}
    p(\vec r,\vec v,t)\propto (\Omega_0/\Omega_\text{eff})^2 \sin^2(\Omega_\text{eff} t/2)
\end{equation}
determines the total excitation rate
\begin{equation}
    \label{eq:exc-prob}
    P_\text{exc}=2\pi \int\int \vec r f(\vec v) n(\vec r,t) p(\vec r,\vec v,t) \text{d}\vec r \text{d}\vec v.
\end{equation}
Note, that the transverse velocity distribution modifies the time-dependent spatial distribution $n(\vec r,t)$. 
This treatment assumes a box pulse of duration $t$ and an effective Rabi frequency
\begin{equation}
    \Omega_\text{eff}= \sqrt{\Omega^2_0(\vec r)+(\vec k_\text{eff}\cdot\vec v)^2},
\end{equation}
where the spatial dependence of the Rabi frequency $\Omega_0(\vec r)$ is given by the radial beam profile. A similar treatment was applied in the case of a recent gravitational wave detection proposal~\cite{Loriani2019}.
\subsection{Wave-front aberrations}\label{sec:WFA}
Matter-light interactions in the atom interferometric cycle are typically subject to the beam's natural wave-front curvature (e.g. of a gaussian beam) and additional imperfections of the laser beam profile~\cite{Schkolnik2015,LouchetChauvet2011} caused by non-ideal optics. While errors due to the initial collimation of large beams (>2\,cm) are negligible, retro-reflection still introduces WFA that lead to a considerable systematic uncertainty. We employ a second order approximation to the deviation from flat wave-fronts for the combined effects of beam and optics.
The resulting spatial dependence of the laser phase fronts imprints a position-dependent phase on the atoms.
Depending on the amplitude and wavelength of the distortion relative to the size of the atomic ensemble, the resulting phase shift may average out, reduce contrast, lead to phase patterns that can be resolved during detection or result in an average phase shift. In state-of-the-art cold atom gravimeters, WFA induce the limiting systematic uncertainty of 30\,nm/s$^2$ to 40\,nm/s$^2$~\cite{Freier2016,Gillot2016}. A more recent analysis of the device in \cite{Gillot2016} evaluated the systematic error to 55\,nm/s$^2$ with an uncertainty of 13\,nm/s$^2$~\cite{Karcher2018}.
We limit our study case to long-scale WFA, assuming a quadratic dependency of the wave-fronts on the transverse position of the atoms, as introduced by a curvature of the retro-reflecting mirror. In this case, the resulting wave-front curvature with radius $R$ couples to the finite velocity spread $\sigma_v$ and induces the phase shift
\begin{equation}
    \label{eq:WFA}
    \sigma_{\phi_\text{WFA}}=\frac{k_\text{eff}}{R} \frac{k_B T_\text{at}}{m_\text{at}}T^2,
\end{equation}
for a Mach-Zehnder (MZ) geometry, a spatial Gaussian density distribution of the ensemble and a Gaussian velocity spread $\sigma_v$~\cite{LouchetChauvet2011}. Here, $k_\text{eff}$ denotes the effective wave vector, $T$ the interrogation time, $k_B$ is the Boltzmann constant and $T_\text{at}$, $m_\text{at}$ refer to the ensemble temperature and atomic mass, respectively.
\subsection{Mean-field effects}\label{subsec:MF-effects}
Mean-field effects arise due to atom-atom interactions in atomic ensembles, scale with growing densities and are an additional source for statistical errors. The mean-field energy reads
\begin{equation}
     E_\text{MF}(r)= g_\text{int}n(r)
\end{equation}
and depends on the local density $n(r)$ of the ensemble and the interaction strength $g_\text{int}^\text{BEC} = 4\pi\hbar^2a_{sc}/m_\text{at}$, where $a_{sc}$ is the s-wave scattering length. For a thermal ensemble, $g_\text{int}^\text{thermal} = 2\,g_\text{int}^\text{BEC}$~\cite{GuryOdelin2002}. Following \cite{Debs2011}, the average mean-field energy for a spherical ensemble of volume $V(t)=4\pi r(t)^3/3$ with $N_\text{at}$ atoms is consequently given by $\langle E_\text{MF}\rangle = g_\text{int} N_\text{at}/V$.
This assumes, however, a uniform density distribution with radius $r$ while thermal and BEC ensembles in fact follow a Gaussian or parabolic distribution, respectively. Hence, we take the average of the mean-field energy by weighting it with the respective density distribution:
\begin{align}
    \langle E_\text{MF}\rangle&=\frac{4 \pi \int_0^\infty dr r^2 n(r) E_\text{MF}(r)}{4 \pi \int_0^\infty dr r^2 n(r)}\notag\\
    &=g_\text{int}\frac{\int_0^\infty dr r^2 n^2(r)}{\int_0^\infty dr r^2 n(r)}\\
    \langle E_\text{MF}^\text{BEC}(t)\rangle&=\frac{15 g_\text{int}^\text{BEC} N_\text{at}^\text{BEC}}{14 \pi \sigma_r(t)^3}\\
    \langle E_\text{MF}^\text{thermal}(t)\rangle&=\frac{g_\text{int}^\text{thermal} N_\text{at}^\text{thermal}}{8 \pi^{3/2} R_\text{TF}(t)^3}
\end{align}
In case of an equal $g_\text{int}$, but unequal atomic density on the two arms of the atom interferometer, a spurious phase shift arises.
Following \cite{Debs2011}, we model the contribution by linking the imbalance in density to the initial beam splitter and neglect effects due to overlap of the two arms or losses.
If the initial beam splitter creates a superposition that deviates by $\sigma_N$ from equal probability in both states, the phase shift
\begin{equation}\label{eq:MF}
\sigma_{\phi_\text{MF}}(t) = \frac{\sigma_N}{\hbar} \int_0^t \langle E_\text{MF}(r(t'))\rangle\,\text{d}t'
\end{equation}
occurs, corresponding to the integral of the differential frequency shift between the arms.
In our assessment, we assume the initial superposition to have equal probabilities of both states on average, but to be affected by white noise with a standard deviation of $\sigma_N$ per cycle.
Without relying on quantum correlations \cite{Brif2020arxiv}, characterization of the beam splitter is limited by quantum projection noise, implying an upper limit of $\sigma_N=1/\sqrt{N_\text{at}}$ per cycle, which we adopt for our analysis.
Consequently, the mean-field-induced phase uncertainty in our model depends on the atom number and implicitly on the time-dependent ensemble size, which allows a trade-off between maximal atom number fluctuation and minimum ensemble size at the first beam splitter.
\subsection{Coriolis effect and gravity gradients}
Two of the most relevant systematic effects are related to the Coriolis force and GGs. The first arises due to the transverse motion of the atoms with respect to the incident beam in combination with Earth's rotation, which forms an effective Sagnac interferometer~\cite{Hogan2008,Mueller2009}. The second is the acceleration uncertainty due to the mass distribution of Earth and the apparatus surrounding the experiment. Both give rise to systematic effects as they couple to the initial kinematic conditions of the ensemble~\cite{Antoine2003}.

In the case of a MZ geometry, the uncertainty in the atom's mean velocity $\delta_v$ and mean position $\delta_r$ couple to GGs $\gamma$ parallel and perpendicular to the sensitive axis. The uncertainty in phase related to gravity gradients is given by
\begin{align}
    \label{eq:GGparallel}
    \delta_{\phi_{v,GG,{\parallel}}}&=k_\text{eff} \gamma_\parallel \delta_v T^3\\
    \delta_{\phi_{r,GG,{\parallel}}}&=k_\text{eff} \gamma_\parallel \delta_r T^2\\
    \label{eq:GGperp}
    \delta_{\phi_{v,GG,{\perp}}}&=\tfrac{14}{3} k_\text{eff} \gamma_\perp \delta_v \Omega T^4\\
    \delta_{\phi_{r,GG,{\perp}}}&=8 k_\text{eff} \gamma_\perp \delta_r \Omega T^3.
\end{align}
Due to low order of magnitude of terms related to the coupling of rotations to transverse GGs, we restrict ourselves to phase uncertainties due to gradients parallel to the sensitive axis. Residual rotations $\Omega$ perpendicular to the sensitive axis couple to the atom's velocity.
The phase uncertainty due to the Coriolis force (subscript C), is given by
\begin{equation}
    \label{eq:Rotations}
        \delta_{\phi_{v,C,{\perp}}}=2 k_\text{eff} \Omega_\perp \delta_v T^2.
\end{equation}
As the density distribution of thermal ensembles is governed by the statistics of a Gaussian density distribution, the uncertainties of the mean position $\delta_r$ and mean velocity $\delta_v$ are related to the spatial $\sigma_r$ and velocity $\sigma_v$ spread, respectively, via
\begin{equation}\label{eq:delta-sigma-relation}
    \delta_{v,r}=\sigma_{v,r}/\sqrt{N_\text{at}\nu_0}.
\end{equation}
The number of atoms $N_\text{at}$ and the number of pre-requisite measurements $\nu_0$ (i.e. the number of times the ensemble is imaged before the actual experiment) equally contribute to the statistical repetition. BECs follow a parabolic density distribution. However, this can be approximated by a Gaussian distribution via \autoref{eq:gauss-parabolic-width}, as explained in the next section.
\subsection{Expansion rate and collimation}
Since expansion rates have sizable effects on the precision and accuracy of atom interferometers, we discuss the possibilities offered by DKC to reduce them in this section. DKC is an established tool to further reduce the effective expansion rate of an ensemble (see~\cite{Corgier2018} and references therein). 
This phase-space-manipulation technique exploits that the free evolution of a gas released from a trap leads to a linear correlation of momentum and radial position of atoms within the ensemble. 
Recapturing the ensemble in a quadratic potential for a well-chosen duration leads to its collimation. 
Preservation of phase space density in the collisionless case requires that this reduction in momentum spread is accompanied by an increased ensemble size and hence necessitates a trade-off between desired expansion rate reduction and required growth in ensemble size. 
For non-interacting gases, i.e. thermal or fermionic gases in all regimes, this relation is captured by Liouville's theorem, 
\begin{equation}
    \sigma_{v_0}/\sigma_{v_f} =  \sigma_{r_f}/\sigma_{r_0},
\end{equation}
which states that the ratio of initial ($\sigma_{v_0}$) to final ($\sigma_{v_f}$) velocity width is inversely proportional to the relative increase in ensemble size ($\sigma_{r_f}/\sigma_{r_0}$). 
A similar relation can be found for interacting degenerate gases - for which the asymptotic final expansion rate is determined by the initial localization through Heisenberg's uncertainty principle and a mean-field contribution~\cite{Loriani2019,Castin1996,Kagan1996} - and for interacting, non-degenerate gases~\cite{GuryOdelin2002,Pedri2003}.
\autoref{fig:DKC_comparison} illustrates a DKC sequence collimating a BEC and a thermal ensemble with typical parameters.
The free expansion of the thermal ensemble is governed by the expansion law
\begin{equation}
    \label{eq:expansion-law}
    \sigma_r (t)=\sqrt{\sigma_{r_0}^2+\sigma_{v_f}^2t^2},
\end{equation}
whereas the BEC dynamics are captured by corresponding scaling laws~\cite{Castin1996}. As an exemplary case, we take a thermal ensemble of $10^9$ $^{87}$Rb atoms at 2\,$\mu$K with a diameter of $2\sigma_r=0.2$\,mm and collimate it down to 80\,nK, such that the required size at lens is $2 \sigma_r(t_\text{DKC})=1$\,mm. For a BEC, we assume an ensemble of $10^6$ $^{87}$Rb atoms and collimate it to 50\,pK in a trap with frequencies of $50\times 2\pi$\,Hz. These are regime-typical parameters for experiments with either thermal ensembles or BECs~\cite{Loriani2019,Mntinga2013,Kovachy2015}. Following~\cite{Loriani2019}, the expansion energies for the thermal atoms and the chemical potential $\mu_\text{BEC}$ of the BEC are obtained from
\begin{align}
    E_\text{thermal}(0)&=\tfrac{3}{2}k_B T_\text{at}(0)\\
    \mu_\text{BEC}(0)&=\tfrac{1}{2} m \omega^2 R_\text{TF}(0)^2\\
    E_\text{thermal}(t_\text{DKC})&=\tfrac{3}{2} k_B T_\text{at}(t_\text{DKC})\\
    E_\text{BEC}(t_\text{DKC})&=\tfrac{1}{2}m\left(\sigma_v(t_\text{DKC})/\sqrt{7}\right)^2
\end{align}
For a better comparison of the two fundamentally dif\-fe\-rent density distributions, the Thomas-Fermi radius $R_\text{TF}$ of the isotropic BEC with parabolic density distribution can be related to a Gaussian spatial width $\sigma_r$ via~\cite{Corgier2018} 
\begin{equation}\label{eq:gauss-parabolic-width}
    R_\text{TF}(t)=\sigma_r(t)\sqrt{7}.
\end{equation}
The resulting characteristics of the collimation sequence for $^{87}$Rb and $^{41}$K ensembles are given in \autoref{tab:DKC}.
\begin{figure}
    \centering
    \begin{tikzpicture}
\begin{groupplot}[
	group style={
	    group size= 2 by 1, 
	    horizontal sep = 0.4cm,
	    xticklabels at=edge bottom
},
	width=1/2*\linewidth+0.1cm,
    height=1.2*(1/2*\linewidth+0.1cm),
    legend style={nodes={scale=0.8, transform shape}}
]

\nextgroupplot[
	xmin=0, xmax=0.55,
	ymin=0, ymax=3.1,	
	ylabel={$2\,\sigma_r(t)$\,(mm)},
	restrict y to domain=0:3.1,
	axis y line*= left,
	xtick={0,0.25,0.5,0.55},
	xticklabels={0,0.25,0.5,}
]

\addplot +[blue,thick,mark=none] table[x index=0, y index=1] {\dataBECdkc};
\addplot +[blue,thick,dashed,mark=none, forget plot] table[x index=0, y index=1] {\dataBECnodkc};
\addplot +[red,thick,mark=none] table[x index=0, y index=1] {\dataCLSdkc};
\addplot +[red,thick,dashed,mark=none, forget plot] table[x index=0, y index=1] {\dataCLSnodkc};

\draw[dotted, thin] ({axis cs:0.025,0}|-{rel axis cs:0,1}) -- ({axis cs:0.025,0}|-{rel axis cs:0,0});

\coordinate (c11) at (rel axis cs:0,1);
\coordinate (c12) at (rel axis cs:1,1);
\coordinate (c21) at (rel axis cs:0,0);
\coordinate (c22) at (rel axis cs:1,0);

\nextgroupplot[
	xmin=1, xmax=10.1,
	ymin=0, ymax=82,	
	restrict y to domain=0:80,
	axis y line*=right,
	xtick={1,2,4,6,8,10},
	xticklabels={,2,4,6,8,10},
	ytick={0,20,40,60,80},
	yticklabels={0,20,40,60,80}
]

\addplot +[blue,thick,mark=none,restrict x to domain=1:10] table[x index=0, y index=1] {\dataBECdkc};
\addplot +[blue,thick,dashed,mark=none, forget plot,restrict x to domain=1:10] table[x index=0, y index=1] {\dataBECnodkc};
\addplot +[red,thick,mark=none,restrict x to domain=1:10] table[x index=0, y index=1] {\dataCLSdkc};
\addplot +[red,thick,dashed,mark=none, forget plot,restrict x to domain=1:10] table[x index=0, y index=1] {\dataCLSnodkc};

\addlegendentry{$^{87}$Rb BEC};
\addlegendentry{$^{87}$Rb thermal};

\coordinate (c13) at (rel axis cs:0,1);
\coordinate (c14) at (rel axis cs:1,1);
\coordinate (c23) at (rel axis cs:0,0);
\coordinate (c24) at (rel axis cs:1,0);

\end{groupplot}

\node[black] at ( $ (c21)!1/2!(c24) +(0,-0.75)$ |-0,-5.5) {t (s)};

\filldraw[color=gray,opacity=0.1] (c12)--(c13)--(c23)--(c22);
\draw[black, dotted,thick] (c12)--(c13);
\draw[black, dotted,thick] (c22)--(c23);

\end{tikzpicture}
    \caption{Size evolution of thermal ensembles (red) and BECs (blue) after release from a trap. A DKC stage is reducing the expansion energies down to 80\,nK and 50\,pK for the thermal and BEC ensembles, respectively (see \autoref{tab:DKC} for the exact parameters). The dashed lines illustrate the expansion in the freely expanding case without collimation. 
    }
   \label{fig:DKC_comparison}
\end{figure}
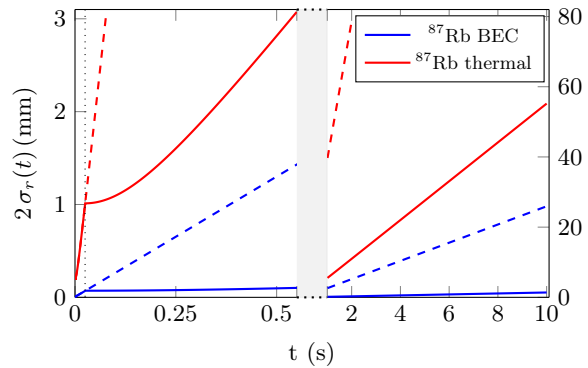
\begin{center}
    \begin{table}[ht]
    \caption{Parameters of the DKC sequence of $^{87}$Rb and $^{41}$K in the thermal and condensed regime.}
    \centering
    \begin{tabular}{@{}l|ll|ll@{}}
\toprule
\multicolumn{1}{c|}{\multirow{2}{*}{Parameter\textbackslash{}Species}} & $^{87}$Rb     & $^{41}$K     & $^{87}$Rb              & $^{41}$K               \\
\multicolumn{1}{c|}{}                                                  & \multicolumn{2}{c|}{thermal} & \multicolumn{2}{c}{BEC}                         \\ \midrule
T$_\text{at}(0)$ / $\mu(0)$ ($\mu$K)                                   & \multicolumn{2}{c|}{2}       & 0.092                  & 0.064                  \\
T$_\text{at}(t_\text{DKC})$ / $\mu(t_\text{DKC})$ (nK)                 & \multicolumn{2}{c|}{80}      & \multicolumn{2}{c}{0.025}                       \\
t$_\text{DKC}$ (ms)                                                    & \multicolumn{2}{c|}{25}      & \multicolumn{1}{c}{26} & \multicolumn{1}{c}{23} \\ \midrule
2$\sigma_r(0)$ (mm)                                                    & 0.2           & 0.3          & 0.010                  & 0.013                  \\
2$\sigma_r(t_\text{DKC})$ (mm)                                         & 1.0           & 1.5          & 0.071                  & 0.079                  \\
$\sigma_v(0)$ (mm/s)                                                   & 14            & 20           & 11                     & 14                     \\
$\sigma_v(t_\text{DKC})$ (mm/s)                                        & 2.770         & 4.130        & 0.183                  & 0.273                  \\
2$\sigma_r(t_\text{DKC}+0.15\,s)$ (mm)                                 & 1.3           & 2.0          & 0.074                  & 0.084                  \\
2$\sigma_r(t_\text{DKC}+0.5\,s)$ (mm)                                  & 3.0           & 4.4          & 0.099                  & 0.130                  \\
2$\sigma_r(t_\text{DKC}+10\,s)$ (mm)                                   & 55            & 83           & 1.385                  & 2.067                  \\ \bottomrule
\end{tabular}
    \label{tab:DKC}
    \end{table}
\end{center}
The illustrated collimation sequence assumes similar free expansion time $t_\text{DKC}$ prior to application of the lens for both regimes. In order to achieve a final expansion behaviour of the thermal ensemble similar to the BEC case, $t_\text{DKC}$ would need to be significantly increased (about two orders of magnitude, depending on the initial temperature), corresponding to a substantially larger ensemble size at the time of the lens~\cite{Loriani2019}.
This is a distinct disadvantage for thermal ensembles, since the DKC technique crucially depends on the harmonicity of the applied lens potential, which has to be verified over the entire spatial extent of the ensemble~\cite{Rudolph2015}.
Application of velocity-selective pulses for temperature reduction in 1D has the disadvantage of atom loss \cite{Kasevich1991}. It is therefore not a promising pathway to reach expansion rates for thermal ensembles comparable to those of BECs. Raman sideband cooling \cite{Hamann1998,Vuleti1998,Estey2015} might be a better alternative for thermal ensembles even if it is limited to about an order of magnitude larger temperatures than what the BEC ensembles could reach.
\subsection{Contrast and detection}
Output states of atom interferometers can be detected by absorption or fluorescence imaging methods.
Which method is appropriate depends on the experimental parameters and the information one wants to acquire.
One main distinction is whether the relative atom numbers in the output port are counted or if atomic density distributions have to be spatially resolved.
Atom number counting is commonly done with fluorescence imaging of the ensembles at the output ports, which relies on the excitation and successive emission of photons by the atoms that are then detected by a simple photo diode or CCD ca\-me\-ra.
The ensemble has to be excited by a laser beam, which means that it has to have a reasonably compact size to be illuminated.
This can usually be achieved with thermal ensembles as well as BECs.
The number of atoms contributing constructively to the signal at the output port is given by the product of the excitation probabilities at all interaction times $t_i$:
\begin{equation}
    \label{eq:contrast}
    C(t_n)=p(t_n)=\prod\limits_{i = 0}^{n}\text{P}_\text{exc}(t_i).
\end{equation}
The contrast C is given by \autoref{eq:contrast} as the convoluted excitation probability for a given phase-space distribution of the ensemble. Inhomogeneous excitation efficiency or phase gradients e.g. caused by GGs may wash out the contrast~\cite{Aguilera2014}.

In experiments employing spatial mapping of the output port wave function for the determination of the phase or analysis of features within the atomic ensemble~\cite{Sugarbaker2013}, good spatial resolution along with a high signal-to-noise ratio and minimal systematic effects during detection are required.
BECs with small spatial spread and expansion rates are thus favored over thermal ensembles to increase the spatial resolution of the CCD picture.
Indeed, the high expansion rates of thermal ensembles may at long times lead to densities challenging for absorption imaging due to decreasing signal per volume.
\section{Comparative performance study}\label{sec:examples}
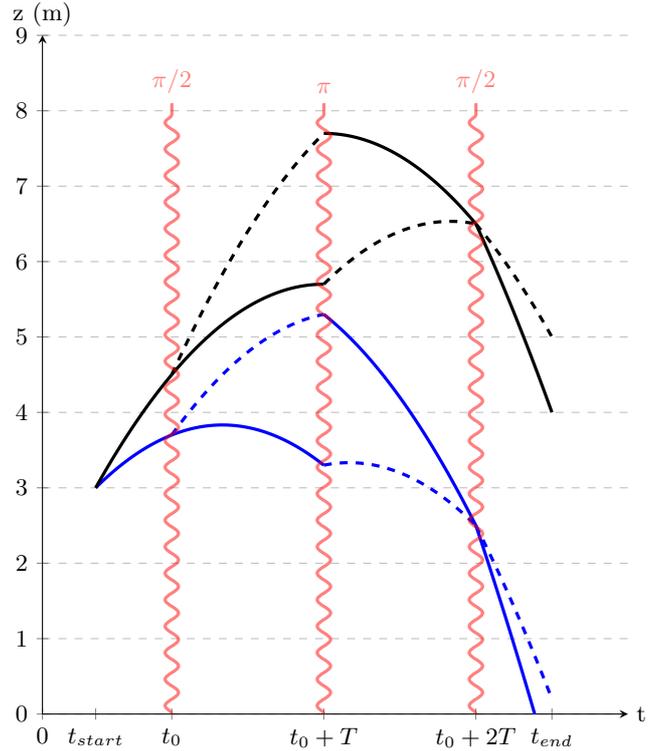
\begin{figure}[ht]
    \centering
    \begin{tikzpicture}
\newcommand{\tmin}{0.7}
\newcommand{\dtmax}{\dtstart}
\newcommand{\dtstart}{1.0}
\newcommand{\T}{2.0}
\newcommand{\zmin}{3.0}
\newcommand{\dvstart}{0.8}
\newcommand{\vstart}{1.0}
\newcommand{\zmax}{9.0}
\newcommand{\g}{0.6}
\newcommand{\keff}{1.0}
\begin{axis}[
    xlabel={t}, xlabel style={at={(1,0)}, anchor=west},
    ylabel={z (m)}, ylabel style={rotate=-90,at={(0,1)}, anchor=south},
    xmin=0.0, xmax=\tmin+\dtstart+2*\T+2*\dtmax,
    ymin=0, ymax=\zmax,
    axis lines = left,
    y=1.cm,
    x=1.cm,
    xtick={0,0.7,1.7,3.7,5.7,6.7},
    xticklabels={$0$,$t_{start}$,$t_0$,$t_0+T$,$t_0+2T$,$t_{end}$},
    ytick={0,1,...,12},
    legend pos=north west,
    ymajorgrids=true,
    grid style=dashed,
    ]
    \addplot[color=blue, samples=50, very thick][domain=\tmin:\tmin+\dtstart]{
        f(
        \zmin
        ,\vstart
        ,(x-\tmin)
        )
    };
    \addplot[color=blue, samples=50, very thick, dashed][domain=\tmin+\dtstart:\tmin+\dtstart+\T]{
        f(
            f(\zmin,\vstart,\dtstart)
            ,\vstart-\g*\dtstart+\keff
            ,x-\tmin-\dtstart
        )
    };
    \addplot[color=blue, samples=50, very thick][domain=\tmin+\dtstart+\T:\tmin+\dtstart+\T+\T]{
        f(
            f(f(\zmin,\vstart,\dtstart),\vstart-\g*\dtstart+\keff,\T)
            ,\vstart-\g*(\dtstart+\T)
            ,x-\tmin-\dtstart-\T
            )
    };
    \addplot[color=blue, samples=50, very thick][domain=\tmin+\dtstart:\tmin+\dtstart+\T]{
        f(
            f(\zmin,\vstart,\dtstart)
            ,\vstart-\g*\dtstart
            ,x-\tmin-\dtstart
        )
    };
    \addplot[color=blue, samples=50, very thick, dashed][domain=\tmin+\dtstart+\T:\tmin+\dtstart+\T+\T]{
        f(
            f(f(\zmin,\vstart,\dtstart),\vstart-\g*\dtstart,\T)
            ,\vstart-\g*(\dtstart+\T)+\keff
            ,x-\tmin-\dtstart-\T
        )
    };
    \addplot[color=blue, samples=50, very thick, dashed][domain=\tmin+\dtstart+2*\T:\tmin+\dtstart+\T+\T+\dtmax]{
        f(
            f(f(f(\zmin,\vstart,\dtstart),\vstart-\g*\dtstart,\T),\vstart-\g*(\dtstart+\T)+\keff,\T)
            ,\vstart-\g*(\dtstart+\T+\T)
            ,x-\tmin-\dtstart-\T-\T
        )
    };
    \addplot[color=blue, samples=50, very thick][domain=\tmin+\dtstart+2*\T:\tmin+\dtstart+\T+\T+\dtmax]{
        f(
            f(f(f(\zmin,\vstart,\dtstart),\vstart-\g*\dtstart,\T),\vstart-\g*(\dtstart+\T)+\keff,\T)
            ,\vstart-\g*(\dtstart+\T+\T)-\keff
            ,x-\tmin-\dtstart-\T-\T
        )
    };    
    \addplot[color=black, samples=50, very thick][domain=\tmin:\tmin+\dtstart]{
        f(
        \zmin
        ,\vstart+\dvstart
        ,(x-\tmin)
        )
    };
    \addplot[color=black, samples=50, very thick, dashed][domain=\tmin+\dtstart:\tmin+\dtstart+\T]{
        f(
            f(\zmin,\vstart+\dvstart,\dtstart)
            ,\vstart+\dvstart-\g*\dtstart+\keff
            ,x-\tmin-\dtstart
        )
    };
    \addplot[color=black, samples=50, very thick][domain=\tmin+\dtstart+\T:\tmin+\dtstart+\T+\T]{
        f(
            f(f(\zmin,\vstart+\dvstart,\dtstart),\vstart+\dvstart-\g*\dtstart+\keff,\T)
            ,\vstart+\dvstart-\g*(\dtstart+\T)
            ,x-\tmin-\dtstart-\T
            )
    };
    \addplot[color=black, samples=50, very thick][domain=\tmin+\dtstart:\tmin+\dtstart+\T]{
        f(
            f(\zmin,\vstart+\dvstart,\dtstart)
            ,\vstart+\dvstart-\g*\dtstart
            ,x-\tmin-\dtstart
        )
    };
    \addplot[color=black, samples=50, very thick, dashed][domain=\tmin+\dtstart+\T:\tmin+\dtstart+\T+\T]{
        f(
            f(f(\zmin,\vstart+\dvstart,\dtstart),\vstart+\dvstart-\g*\dtstart,\T)
            ,\vstart+\dvstart-\g*(\dtstart+\T)+\keff
            ,x-\tmin-\dtstart-\T
        )
    };
    \addplot[color=black, samples=50, very thick][domain=\tmin+\dtstart+2*\T:\tmin+\dtstart+\T+\T+\dtmax]{
        f(
            f(f(f(\zmin,\vstart+\dvstart,\dtstart),\vstart+\dvstart-\g*\dtstart,\T),\vstart+\dvstart-\g*(\dtstart+\T)+\keff,\T)
            ,\vstart+\dvstart-\g*(\dtstart+\T+\T)-\keff
            ,x-\tmin-\dtstart-\T-\T
        )
    };  
    \addplot[color=black, samples=50, very thick, dashed][domain=\tmin+\dtstart+2*\T:\tmin+\dtstart+\T+\T+\dtmax]{
        f(
            f(f(f(\zmin,\vstart+\dvstart,\dtstart),\vstart+\dvstart-\g*\dtstart,\T),\vstart+\dvstart-\g*(\dtstart+\T)+\keff,\T)
            ,\vstart+\dvstart-\g*(\dtstart+\T+\T)
            ,x-\tmin-\dtstart-\T-\T
        )
    };  
\end{axis}

    \draw[decorate, decoration = snake, very thick, red, opacity=0.5]
    ({\tmin + \dtstart},0) to ({\tmin + \dtstart},{.9*\zmax}) node[above] {$\pi/2$};
    \draw[decorate, decoration = snake, very thick, red, opacity=0.5,] 
    ({\tmin + \dtstart+\T},0) to ({\tmin + \dtstart + \T},{.9*\zmax}) node[above] {$\pi$};
    \draw[decorate, decoration = snake, very thick, red, opacity=0.5,] 
    ({\tmin + \dtstart + 2*\T},0) to ({\tmin + \dtstart + 2*\T},{.9*\zmax}) node[above] {$\pi/2$};
\end{tikzpicture}
    \caption{The geometry for a gradiometer consists of two MZ sequences that are operated simultaneously (black and blue lines). Each MZ sequence is formed by three consecutive light pulses that constitute a beam splitter (first $\pi/2$-pulse), a mirror ($\pi$-pulse) and a merging beam splitter after $2T$. The dashed lines indicate a change of the internal state due to a momentum transfer by the atom-light interaction. 
    A single MZ sequence corresponds to a gravimeter sequence to measure the coupling constant $g$ of the gravitational field to one species.
    Two MZ sequences operated with two different species allow for a determination of $\eta$. In this case the differential velocity at $t=t_\text{start}$ of the MZ sequences is set such that they spatially overlap to eliminate the gravity gradient, but remain sensitive to a potentially species specific differential acceleration.}
    \label{fig:AI-geometry-scheme}
\end{figure}

Based on the discussion of phase shifts and performance-limiting effects in the previous section, we now elaborate on three study cases in which we compare quantum degenerate ensembles to thermal sources.
In highly dynamical environments, such as inertial sensing and for navigation purposes, thermal sources may be beneficial since they typically feature more atoms and shorter cycle times, which decreases both, shot noise and integration time. However, a trade-off has to be found for every particular situation due to their relatively high expansion rates and spatial extension.
The three specific examples selected for the comparison consist of inertial sensors that could operate beyond state-of-the-art in the near future: \textit{(i)} A ground-based gravimeter with a relative uncertainty of $\Delta g/g=10^{-9}$, \textit{(ii)} a space gradiometer with a $2.5$\,mE resolution \cite{Trimeche2019} and \textit{(iii)} a WEP-test with an uncertainty of $2\times10^{-15}$ in the determination of the Eötvös ratio \cite{Aguilera2014}. The interferometer geometries are illustrated in \autoref{fig:AI-geometry-scheme}.

In order to evaluate the performance of every regime, we will take typical parameters for thermal ensembles and BECs and assess their performance in each of the three experiments. Details of the ensemble sizes and velocity distributions for both regimes are given in \autoref{tab:DKC} at key times of the interferometric sequences.
We assume a Gaussian beam of 3\,cm 1/$e^2$-radius with Rabi-frequencies of $\pi/(25\times 10^{-6})$\,Hz (gravimeter and WEP-test) and $\pi/(55\times 10^{-6})$ Hz (gradiometer) for second order beam splitting processes, GGs of $10^{-6}$\,$s^{-2}$ and spurious rotations on the order of 1\,$\mu$rad/s due to imperfect rotation compensation of the mirror and limited attitude control of the satellite for the two space-borne missions. The wave-front curvature is assumed to be 2.3\,km (gravimeter), 5.6\,km (gradiometer) and 250\,km (WEP-test). 

As illustrated in the previous sections, systematic effects linked to GGs and the Coriolis force are connected to the uncertainty of the mean position $\delta_r$ and velocity $\delta_v$ at the beginning of the interferometry sequence. The number of measurements $\nu_0$ required for their characterization (see \autoref{eq:delta-sigma-relation}) is determined for each application such that the largest systematic phase uncertainty related to GGs or rotations (\autoref{eq:GGparallel} and \autoref{eq:Rotations}) is below the target uncertainty chosen for the respective measurement.

The number of prerequisite experiments sufficient to suppress the systematic effects below the target uncertainty may differ for the BEC and the thermal case. In our cases, thermal ensembles have three orders of magnitude more atoms and are about 15-20 times larger than BECs after the DKC. Hence, the minimal number of characterization measurements $\nu_0$ is reduced by a factor of 2 to 5 compared to the BEC case.
For the sake of comparability, we choose to compute all systematic effects with $\nu_0^\text{BEC}$. This enables an evaluation of the performance with a fixed set of parameters.

As the systematic phase uncertainty due to WFA depends on the velocity width and not the mean velocity (see \autoref{eq:WFA}), its magnitude does not depend on the number verification measurements. Thus, it can neither be integrated down nor reduced by prerequisite measurements, conversely to the GGs. Its value is completely predetermined by the ensemble's expansion rate.

In order to adapt statistical error contributions such as shot noise and mean-field effects to the desired precision of every type of measurement, we calculate the minimum number of iterations n$_\text{cycle}$ until the target uncertainty is reached. The integrated (denoted by subscript $i$) shot noise is given by 
\begin{equation}
    \label{eq:shot-noise}
    \sigma_{\phi_\text{SN},i}=\sigma_{\phi_\text{SN}}/\sqrt{n_\text{cycle}}=1/\sqrt{N_\text{at} n_\text{cycle}}C,
\end{equation}
where the contrast $C$ is given by \autoref{eq:contrast}, i.e. the convolved excitation probability. Non-perfect contrast $(C<1)$ increases the shot noise as the number of atoms constituting the statistical sample is reduced.

Mean-field effects contribute a statistical phase uncertainty expressed by \autoref{eq:MF} where the ensemble expansion over time is taken into account. This effect integrates down with the number of experiments n$_\text{cycle}$ following
\begin{equation}
    \sigma_{\phi_\text{MF},i}=\sigma_{\phi_\text{MF}}/\sqrt{n_\text{cycle}}.
\end{equation}

Once the number of cycles is determined by the desired performance, the associated integration time t$_\text{int}$ depends on the preparation time t$_\text{prep}$ and the interrogation time 2T:
\begin{equation}
    \label{eq:integration-time}
    \text{t}_\text{int}=\left(\text{t}_\text{prep}+2T\right)\text{n}_\text{cycle}=\text{t}_\text{cycle}\text{n}_\text{cycle}.
\end{equation}

For a straightforward comparison of the performance differences between BEC and thermal ensemble, the integration time is also chosen to be the same for both regimes, initially determined by the number of cycles needed to suppress the statistical effects of the BEC below the target uncertainty. Since thermal sources can be generated within a shorter preparation time, more cycles can be performed during the same integration time according to
\begin{equation}
    n_\text{cycle}^\text{thermal}=n_\text{cycle}^\text{BEC}\times t_\text{cycle}^\text{BEC}/t_\text{cycle}^\text{thermal}.
\end{equation}

In order to estimate the various uncertainties, ensemble properties as spatial and velocity spreads are computed at each atom-light interaction pulse, to take into account the spatial and velocity selectivity of the pulses applying \autoref{eq:exc-prob} and \autoref{eq:contrast}.
The modified spatial and velocity spreads are the evaluation input for the mean-field effects, the WFA and the estimation of the contrast according to the formulae given in \autoref{sec:physics}.
The results of this study are summarized in \autoref{tab:numbers} where the phase uncertainties are normalized as fractional phases $\phi/k_\text{eff}gT^2$ for the gravimeter and the WEP test. In the gradiometer case, the orders of magnitude are given in units of $\phi/k_\text{eff}DT^2=\Gamma$ for a baseline D. Here, $\phi_\text{target}$ is the upper limit for any systematic or statistical phase uncertainty.

Detailed results for the three science cases are presented in the consecutive sections.

\begin{center}
\begin{table*}[ht]
    \caption{Estimation of statistical and systematic uncertainties for three scenarios: a lab-based $^{87}$Rb gravimeter~\cite{LouchetChauvet2011}, a space-borne $^{87}$Rb gradiometer~\cite{Trimeche2019} and a satellite $^{87}$Rb/$^{41}$K WEP-test analogous to the STE-QUEST mission~\cite{Aguilera2014}. The phase uncertainties are given as fractions $\Delta a/g=\delta_\phi/k_\text{eff}gT^2$ (gravimeter and WEP-test) and $\Delta \Gamma=\delta_\phi/k_\text{eff}DT^2$ (space gradiometer). The expansion sequence over the course of the atom interferometer is calculated in \autoref{tab:DKC}. Systematic effects are denoted by $\delta$, while statistical effects are denoted by $\sigma$ and calculated after integrating over a number $n_\text{cycle}$ of experiments with N$_\text{at}$ atoms in each cycle. Gravity gradients are abbreviated with GG, the Coriolis effect with C, wave-front aberrations by WFA, shot noise with SN and mean-field effects by MF.}
    \centering
\begin{tabularx}{0.95\linewidth}{@{}l|ll|ll|ll@{}}
\toprule
\multicolumn{1}{c|}{\multirow{2}{*}{Parameter\textbackslash{}Case}} & \multicolumn{2}{c|}{Gravimeter}             & \multicolumn{2}{c|}{Gradiometer}                                   & \multicolumn{2}{c}{WEP-test}                              \\ \cmidrule(l){2-7} 
\multicolumn{1}{c|}{}                                               & thermal              & BEC                  & thermal                          & BEC                             & thermal                          & BEC                    \\ \midrule
N$_\text{at}$ (initial)                                             & $1\times 10^9$       & $1\times 10^6$       & $1\times 10^9$                   & $1\times 10^6$                  & $1\times 10^9$                   & $1\times 10^6$         \\
T$_\text{at}$ (K)                                                   & $80\times 10^{-9}$   & $50\times 10^{-12}$  & $80\times 10^{-9}$               & $50\times 10^{-12}$             & $80\times 10^{-9}$               & $50\times 10^{-12}$    \\
P$_\text{exc}$                                                      & 0.57                 & 0.99                 & 0.33                             & 0.97                            & 0.43                             & 0.99                   \\ \midrule
t$_\text{int}$ (s); n$_\text{cycle}$                                & 3.36; 7              & 3.45; 3              & 86400; 72000                     & 86400; 4320                     & $2\times 10^7$; $2.4\times 10^7$ & $2\times 10^7$; $10^6$ \\
$\nu_0$                                                             & \multicolumn{2}{c|}{1}                      & \multicolumn{2}{c|}{50}                                            & \multicolumn{2}{c}{$10^6$}                                \\
2T (s)                                                              & \multicolumn{2}{c|}{150$\times 10^{-3}$}    & 0.5                              & 10                              & 0.5                              & 10                     \\
$\mathcal{O}(\delta_{\phi_\text{target}})$                          & \multicolumn{2}{c|}{$\Delta g/g=10^{-9}$}   & \multicolumn{2}{c|}{$\Gamma=2.5$ mE$=2.5\times 10^{-12}$ s$^{-2}$} & \multicolumn{2}{c}{$\eta=2\times 10^{-15}$}               \\
$\mathcal{O}(\sigma_{\phi_\text{SN}})$                              & $1.2\times 10^{-11}$ & $3.3\times 10^{-10}$ & $5.0\times 10^{-13}$ s$^{-2}$    & $5.5\times 10^{-14}$ s$^{-2}$   & $1.2\times 10^{-15}$             & $2.1\times 10^{-16}$   \\ \midrule
$\mathcal{O}(\delta_{\phi_\text{GG}})$                              & $4.9\times 10^{-15}$ & $1.1\times 10^{-14}$ & $2.4\times 10^{-14}$ s$^{-2}$    & $9.9\times 10^{-13}$ s$^{-2}$   & $1.1\times 10^{-17}$             & $4.8\times 10^{-16}$   \\
$\mathcal{O}(\delta_{\phi_\text{C}})$                               & $1.8\times 10^{-14}$ & $3.7\times 10^{-14}$ & $7.0\times 10^{-14}$ s$^{-2}$    & $1.5\times 10^{-13}$ s$^{-2}$   & $3.7\times 10^{-17}$             & $7.6\times 10^{-17}$   \\
$\mathcal{O}(\delta_{\phi_\text{WFA}})$                             & $3.4\times 10^{-10}$ & $2.1\times 10^{-13}$ & $1.0\times 10^{-12}$ s$^{-2}$    & $4.4\times 10^{-15}$ s$^{-2}$   & $1.3\times 10^{-12}$             & $8.0\times 10^{-16}$   \\
$\mathcal{O}(\sigma_{\phi_\text{MF}})$                              & $1.3\times 10^{-11}$ & $9.2\times 10^{-10}$ & $4.7\times 10^{-13}$ s$^{-2}$    & $5.4\times 10^{-13}$ s$^{-2}$   & $1.1\times 10^{-15}$             & $1.8\times 10^{-15}$   \\ \bottomrule
\end{tabularx}
    \label{tab:numbers}
\end{table*}
\end{center}

\subsection{Gravimeter}
We start by comparing two ground-based $^{87}$Rb gravimeters operated with thermal atoms or BECs, the source characteristics of which are similar to the ones reported in~\cite{LouchetChauvet2011} and~\cite{Albers2020,Rudolph2015}. In both cases, the interrogation time 2T equals 150\,ms.

The first column in \autoref{tab:numbers} shows the magnitudes of the different performance limiting effects discussed in the previous section in units of $\Delta g/g$. The scenario targets an uncertainty of 1 $\mu$Gal=$10^{-8}$ m/s$^2$, corresponding to a fractional phase uncertainty of $\delta_{\phi_\text{target}}/k_\text{eff}gT^2=10^{-9}$. 

A thermal ensemble with $10^9$ $^{87}$Rb atoms would have a diameter of $2\sigma_r(t_\text{DKC})=1$\,mm after the DKC pulse. The velocity spread at the lens is $2.7$\,mm/s. The convolved excitation efficiency at the last beam splitter is 57\%. 

Adopting a BEC-source as in~\cite{Rudolph2015}, it is reasonable to assume a collimation of the ensemble to an effective temperature of 50\,pK for $10^6$ atoms. The convolved excitation efficiency at the last beam splitter is 99\% for an ensemble diameter of $2R_\text{TF}(t_\text{DKC})=0.19$\,mm and a velocity spread of 183\,$\mu$\,m/s. Although the order of magnitude of the initial atom number is three times larger for the thermal atoms, the shot noise is only one order of magnitude below the one of the BEC due to the reduced contrast.

Theoretically, the target uncertainty of $\Delta g/g=10^{-9}$ is reached in both cases after only one verification shot and integration over seven (thermal, full cycle time 0.48\,s) or three (BEC, full cycle time 1.15\,s) experimental cycles corresponding to a few seconds of integration time. 
All statistical and systematic effects are below the target uncertainty hinting towards the possibility to use either of the source concepts.

However, if a better performance of the gravimeter is sought for, the first limit to tackle would be the mean-field effects at $9.2\times10^{-10}$ in the BEC case and wave-front distortions at $3.4\times10^{-10}$ in the thermal one. 
Mean-field effects are a statistical phenomenon, hence one can integrate the phase uncertainty for the BEC-case down by increasing the number of experiments. 

As mentioned, WFA are not a statistical but a systematic phenomenon and cannot be integrated down by adding verification shots. Therefore thermal sources are limited by WFA to the $10^{-10}$ level, whereas
the BECs could improve on the accuracy up to the $10^{-13}$ level.

\subsection{Gradiometer}
Here, we are address a satellite gradiometer as proposed in \cite{Trimeche2019}. It features a baseline of $D=0.5$\,m separating the two interferometers, an interrogation time of 2T=10\,s (full cycle time of 20\,s) and a targeted uncertainty of 2.5\,mE, clearly beyond the current state of the art. We adapt the interferometry time from 2T=10\,s to 2T=0.5\,s to constrain the ensemble size at a detectable level in the thermal case.

The center column of \autoref{tab:numbers} displays the order of magnitude of uncertainties in the gradient determination related to the different effects. Here, the numbers are given as spurious gradients in units of $\Gamma$ by dividing each phase uncertainty by $k_\text{eff}DT^2$.

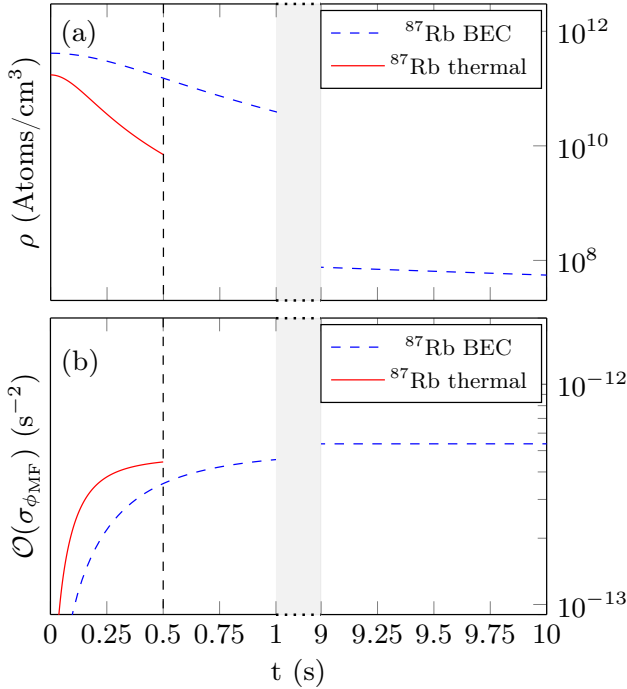
\begin{figure}[h]
    \centering
    \resizebox{\linewidth}{!}{
        \begin{tikzpicture}
    \begin{groupplot}[group style={group size= 2 by 2
    								,vertical sep=0.2cm
    								,horizontal sep=0.5cm
    								,xticklabels at=edge bottom
    								}
    					,width=1/2*\linewidth-0.3cm
    					,height=1.2*(1/2*\linewidth-0.3cm)
    					,legend style={nodes={scale=0.8, transform shape}}	,xtick={0,0.25,0.5,0.75,1,9.,9.25,9.5,9.75,10}
    					]
        \nextgroupplot[xmin=0
        				,xmax=1.0
        				,ymin=2*10^7
        				,ymax=3*10^12
        				,ymode=log
        				,axis y line*=left
        				,ytick=\empty
        				,ylabel={$\rho$ (Atoms/cm$^3$)}]
        				]
                \addplot[blue,dashed,domain=0:1] 
                		table [x expr=\thisrowno{0},y expr=\thisrowno{1}*10^-6] \DensityDataBEC;
                \addplot[red,domain=0:1]
                		table [x expr=\thisrowno{0},y expr=\thisrowno{1}*10^-6] \DensityDataTHE;
                \draw[dashed, thin] ({axis cs:0.5,0}|-{rel axis cs:0,1}) -- ({axis cs:0.5,0}|-{rel axis cs:0,0});
                \coordinate (top11) at (rel axis cs:1,1);
                \coordinate (bot11) at (rel axis cs:1,0);
                \node [text width=2em,anchor=north west] at (rel axis cs: -0.05,1.05)
                {\subcaption{\label{fig:density-plot}}};
        \nextgroupplot[xmin=9.
        				,xmax=10.
        				,ymin=2*10^7
        				,ymax=3*10^12
        				,ylabel={}
        				,ymode=log
        				,axis y line*=right
        				]
                \addplot[blue,dashed,domain=9:10]
                		table [x expr=\thisrowno{0},y expr=\thisrowno{1}*10^-6] \DensityDataBEC;
                \addlegendentry{$^{87}$Rb BEC}
                \addplot[red,domain=9:10]
                		table [x expr=\thisrowno{0},y expr=\thisrowno{1}*10^-6] \DensityDataTHE;
                \addlegendentry{$^{87}$Rb thermal}
                \coordinate (top12) at (rel axis cs:0,1);
                \coordinate (bot12) at (rel axis cs:0,0);
        \nextgroupplot[xmin=0
        				,xmax=1.
        				,ymin=10^-13
        				,ymax=2*10^-12
        				,ymode=log
        				,ylabel={$\mathcal{O}(\sigma_{\phi_\text{MF}})$ (s$^{-2}$)}
        				,axis y line*=left
        				,ytick=\empty
        				]
        		\node [text width=1em,anchor=north west] at (rel axis cs: 0,1)
                {\subcaption{\label{fig:sigma-MF-plot}}};
                \addplot[blue,dashed,domain=0:1] table \ErrorDataBEC;
                \addplot[red,domain=0:1] table \ErrorDataTHE;              
                \draw[dashed, thin] ({axis cs:0.5,0}|-{rel axis cs:0,1}) -- ({axis cs:0.5,0}|-{rel axis cs:0,0});
                \coordinate (top21) at (rel axis cs:1,1);
                \coordinate (bot21) at (rel axis cs:1,0);
        \nextgroupplot[xmin=9.
        				,xmax=10
        				,ymin=0.9*10^-13
        				,ymax=2*10^-12
        				,ymode=log
        				,axis y line*=right
        				]
                \addplot[blue,dashed] table \ErrorDataBEC;
                \addlegendentry{$^{87}$Rb BEC}
                \addplot[red] table \ErrorDataTHE;
                \addlegendentry{$^{87}$Rb thermal}                
                \draw[dashed, thin] ({axis cs:0.5,0}|-{rel axis cs:0,1}) -- ({axis cs:0.5,0}|-{rel axis cs:0,0});
                \coordinate (top22) at (rel axis cs:0,1);
                \coordinate (bot22) at (rel axis cs:0,0);
    \end{groupplot}
    \filldraw[gray,opacity=0.1] (top11)--(top12)--(bot12)--(bot11);
    \filldraw[gray,opacity=0.1] (top21)--(top22)--(bot22)--(bot21);
    \draw[black, dotted,thick] (top11)--(top12);
    \draw[black, dotted,thick] (bot11)--(bot12);
    \draw[black, dotted,thick] (top21)--(top22);
    \draw[black, dotted,thick] (bot21)--(bot22);
    \node[black] at ( $ (bot21)!1/2!(bot22) +(0,-0.65)$ |-0,-5.5) {t (s)};
\end{tikzpicture}
        }
    \caption{Atomic densities and time averaged mean-field statistical uncertainty for the space-gradiometer scenario described in \autoref{tab:numbers}. (a) Time-dependent density $\rho$ of the ensembles during the interferometer time based on the DKC sequence described in \autoref{tab:DKC}. (b) Fractional statistical phase uncertainty $\mathcal{O}(\sigma_{\phi_\text{MF}})$ due to mean-field effects according to equation \autoref{eq:MF} integrated over a number of $n_\text{cycle}$ experiments. }
    \label{fig:MF-effects}
\end{figure}

The phase uncertainties due to GGs, the Coriolis force and mean-field effects receive contributions from both individual interferometers. Through \autoref{eq:delta-sigma-relation}, the initial spatial and velocity spreads $\sigma_{r,v;1,2}$ of interferometer 1,2 enter the systematic uncertainties given in \autoref{eq:GGparallel} and \autoref{eq:Rotations} as 
\begin{equation}
    \label{eq:mean-square-spreads}
    \sigma_{r,v}=\sqrt{\sigma_{r,v;1}^2+\sigma_{r,v;2}^2},
\end{equation}
supposing uncorrelated source noise. For initial spreads with the same width $\sigma_{r,v;1}=\sigma_{r,v;2}$, this yields a factor of $\sqrt{2}$, which results in an integration behaviour during the verification measurements of
\begin{equation}
    \delta_{\phi,i}=\delta_{\phi}\sqrt{2}/\sqrt{\nu_0},
\end{equation}
in the case of GGs and Coriolis effect.
The same holds true for the integrated shot noise, which is increased by a factor of $\sqrt{2}$ compared to \autoref{eq:shot-noise}:
\begin{equation}
    \sigma_{\phi_\text{SN},i}=\sqrt{2}/\sqrt{N_\text{at} n_\text{cycle}}C,
\end{equation}
and for the mean-field effects, which are uncorrelated between the two branches of the interferometer:
\begin{equation}
    \sigma_{\phi_\text{MF},i}=\sqrt{2}\sigma_{\phi_\text{MF}}/\sqrt{n_\text{cycle}}.
\end{equation}
Interestingly, following our treatment in \autoref{subsec:MF-effects}, we find that due to the different expansion behaviour and a higher atom number, the mean-field effects of the thermal ensemble are comparable, in magnitude, to that of the BEC on the time scales we are investigating (see \autoref{tab:numbers} and \autoref{fig:MF-effects}).

Assuming the same velocity spread for the two ensembles in the two interferometers and - as for the gravimeter case - the simplification of a constant curvature, the differential phase shift vanishes. Here, we drop this simplification and consider the propagation of a Gaussian beam which leads to a local dependency of the curvature, and thus to a non-vanishing phase shift in the differential signal. Assuming a residual radius of curvature of the retro-reflection mirror of 4\,km and a propagating the laser beam as outlined in \cite{Trimeche2019} introduces the differential phase shift as reported in~\autoref{tab:numbers}.
 
To reach the uncertainty goal of 2.5\,mE, an integration time on the order of 1 day is required in both cases. Albeit the high expansion rate of the thermal ensemble is accounted for with a shorter interrogation time of 0.5\,s (full cycle time of 1.2\,s), the shot noise uncertainty in the thermal case exceeds the one of the BEC case as the contrast drops to 33\% (97\% for the BEC after 10\,s).
Again, the WFA would hinder any further attempts for a better performance below the $10^{-12}\,s^{-2}$ level. All other systematic effects are complying with the performance required from this sensor and can be reduced by an increased number of verification shots. 

Being limited by WFA at the level of $4.4\times10^{-15}$, the BEC clearly leaves more room for improvement compared to thermal sources, which are bound three orders of magnitude above. Therefore, using a BEC source in this scenario is advantageous since it does neither suffer from a worse integration time nor a larger mean-field effect, yet significantly extends the desired sensitivity range compared to the thermal source.

\subsection{WEP-test}\label{sec:WEP-test}
The WEP-test example is based on the parameters of the STE-QUEST satellite mission proposed in~\cite{Aguilera2014}. We here assume the test pair $^{41}$K and $^{87}$Rb. It aims at an Eötvös ratio $\eta$ determined with an uncertainty of $2\times10^{-15}$. As for the gradiometer, we compare a thermal ensemble at 0.5\,s of interrogation time (full cycle time of 0.83\,s) with a BEC at 10\,s (full cycle time of 20\,s) of interrogation time to ensure non-vanishing contrast and technically manageable ensemble sizes in both cases. In the last column of \autoref{tab:numbers}, the results of the comparison are displayed.

For the determination of the systematic effects one has to take into account the differences of the two atomic species such as the different expansion rates, atomic masses and initial conditions like spatial spread. The two different species-specific excitation rates are also evaluated and the minimum given in \autoref{tab:numbers}.

The effects of GGs and Coriolis force do not scale with $\sqrt{2}$ in this case, but rather with the mean square of two uncorrelated spreads (see \autoref{eq:mean-square-spreads}).
With two different species propagating in the arms, the mean-field effects are calculated as the mean-square sum of the individual mean-field effects of $^{87}$Rb and $^{41}$K as in \autoref{eq:MF}.
As for the gradiometer, one benefits from the differential suppression of WFA when matching the expansion rates of the ensembles~\cite{Aguilera2014}. As two different sources and two different species are used for the production of the ensembles, the systematic differential phase uncertainty is given by
\begin{equation}
    \delta_{\phi_\text{WFA}}= \frac{k_\text{eff}}{R} k_B \left(\frac{T_\text{at,K}}{m_\text{at,K}}-\frac{T_\text{at,Rb}}{m_\text{at,Rb}}\right)T^2,
\end{equation}
analogous to \autoref{eq:WFA}. 
By experimentally matching the velocity spreads of the two ensembles to the 20\% level in both arms, the WFA are suppressed by a factor of 3.

In the BEC case and in order for the statistical effects to be consistent with the mission performance goal, $10^6$ experimental cycles $n_\text{cycle}$ are needed. This leads to a full mission time on the order of five years in case of a highly elliptical orbit operation.
To reach the targeted uncertainty, an additional $10^6$ verification measurements $\nu_0$ are necessary. These measurements are included in the total mission time as they are performed on the transition between perigee and apogee parts of the orbit not dedicated to science measurements~\cite{Aguilera2014}. The contrast at the end of the sequence is at 99\% for both, $^{87}$Rb and $^{41}$K, and for the chosen set of parameters it is feasible to achieve the missions goals with condensed ensembles.

With a thermal ensemble, one would need an integration time on the order of $10^8$ s to suppress mean-field effects, and $10^6$ verification shots to estimate the phase uncertainty due to GGs and the Coriolis effect to a sufficient level. Moreover, even for the reduced interrogation time of 0.5 s, the contrast is at 43\% and the shot noise is almost 6 times larger than in the BEC case.

We conclude that - for the chosen set of parameters - it is possible to achieve the mission goals with condensed ensembles. The thermal case is, however, limited to the $10^{-12}$ level due to the WFA effects stemming from the large expansion rates of the thermal ensembles~\cite{Karcher2018}. In order to reach a reduced velocity spread with a thermal ensemble comparable to that of a BEC, the DKC would be required to handle ensembles with a diameter on the order of 0.5\,m after a pre-lens free expansion time of several seconds. As for the space-gradiometer, this is unpractical for obvious manipulation and excitation reasons. In the condensed regime, a simultaneous collimation of the dual-source was recently considered in reference~\cite{Corgier2020} and shown to be feasible.
\section{Discussion and conclusion}
In this paper, the current limits for state-of-the-art precision experiments with atom interferometry were analyzed. A particular emphasis was put on the comparison of the statistical and systematic uncertainties between condensed and thermal ensembles. Three detailed study cases of a lab-based gravimeter, a space gradiometer and a satellite WEP-test were chosen to illustrate the limits of each regime.

Thermal sources benefit from a shorter cycle time and larger atom numbers compatible with experiments where moderate scale factors suffice or rapid readouts are required. This is, however, beneficial at short interferometry times only. When going beyond state-of-the-art, i.e. from drift times of a fraction of a second to a few seconds, this advantage is lost.
In our study cases, the scenarios utilizing BECs show - at equal integration time - the same shot-noise level and the magnitude of the mean-field effects is comparable to that of thermal sources. 
Moreover, the condensed sources benefit from a very large contrast (close to 1) when compared to their thermal counterparts. 
More dramatically, the WFA set an ultimate limit for thermal ensembles that would not be compatible with long interrogation times, which are required for advanced scenarios. 
For BEC ensembles, their compact sizes make this limit at least three orders of magnitude lower, highlighting their potential in the field of metrology.

Small scale distortions (few $\mu m$) of the optical beams~\cite{Bade2018}, not considered in this article, can hint to a disadvantage for the BEC samples by means of averaging effects for WFA.
However, the flexibility in tuning their initial size~\cite{Corgier2020} mitigates this effect and could bring them to starting sizes similar to thermal ensembles if necessary. 
This engineering of the BEC size allows for a distinct analysis of WFA with long and short periodicity~\cite{LouchetChauvet2011,Karcher2018}. 
This might be especially relevant on short time scales, i.e. for very small ensemble sizes. Their subsequent expansion could still be limited to a few mm thanks to the DKC technique. 
In consequence, a trade-off between the size-stretch-induced phase uncertainties, e.g. to balance the level of GGs or Coriolis systematics versus WFA effects is required. 
This appears to be feasible, especially if one considers gravity gradient compensation schemes as the one in \cite{Loriani2020,Trimeche2019,Roura2017}.

Other considerations that are not reflected by our study would further consolidate the BEC choice. Indeed, we optimistically anticipate here that thermal ensembles could be collimated to the 80\,nK level and that the same level of efficiency in preparing, transporting and engineering of their the quantum states can be achieved as for BECs.
As a conclusion, thermal and BEC sources could equally be employed in relatively short interferometry times (a few hundred ms) for the same performance. With respect to longer times, BEC sources are clearly more advantageous since size-related systematic effects are several orders of magnitude smaller than those of thermal ensembles.
\section{Authors contributions}
All the authors were involved in the preparation of the manuscript.
All the authors have read and approved the final manuscript.
\section{Acknowledgements}
The authors would like to thank Dennis Schlippert and Waldemar Herr for constructive criticism of the manuscript. This work is supported by the German Space Agency (DLR) with funds provided by the Federal Ministry for Economic Affairs and Energy (BMWi) due to an enactment of the German Bundestag under Grant Nos. 50WM\-1861 and 50WM\-2060, by ``Nieders\"achsisches Vorab" through the ``Quantum- and Nano-Metrology (QUANOMET)" initiative within the project QT3, through the Deutsche For\-schungs\-ge\-mein\-schaft (DFG, German Research Foundation) under Germany's Excellence Strategy – EXC 2123 QuantumFrontiers, Project-ID 390837967 and under the CRC 1227 (DQmat) within Projecs No. A05 and No. B07, and through ”Förderung von Wissenschaft und Technik in For\-schung und Lehre” for the initial funding of research in the new DLR Institute (DLR-SI). We also acknowledge support by the QUEST-LFS, the Verein Deutscher Ingenieure (VDI) with funds provided by the Federal Ministry of Education and Research (BMBF) under Grant No. VDI 13N14838 (TAIOL).
SL acknowledges the support of the IP@Leibniz program of the Leibniz University of Hanover for travel grants supporting his stays in France. NG acknowledges mobility support from the Q-SENSE project, which has received funding from the European Union's Horizon 2020 Research and Innovation Staff Exchange (RISE) Horizon 2020 program under Grant Agreement Number 691156.

\end{document}